\documentstyle[psfig,prd,aps]{revtex}

\def\be{\begin{equation}}
\def\ee{\end{equation}}
\def\bea{\begin{eqnarray}}
\def\eea{\end{eqnarray}}

\begin{document}
\draft
\preprint{TAN-FNT-98-03}

\title{Casimir corrections in the bound state soliton model}

\author{Norberto N. SCOCCOLA$^{a}$ and Hans WALLISER$^b$}

\address{
$^a$ Physics Department, Comisi\'on Nacional de Energ\'{\i}a At\'omica,\\
	  Av.Libertador 8250, (1429) Buenos Aires, Argentina.\\
$^b$ Departamento de F\'{\i}sica, Universidade de Coimbra,\\ 
          P-3000 Coimbra, Portugal.\\}

\date{May 1998}

\maketitle
\begin{abstract}
\noindent
We consider the one-loop corrections to the $SU(3)$ skyrmion mass within
the bound state soliton approach. We show that the standard $SU(3)$ 
renormalization 
scheme is not appropiate within this framework and propose to use an 
alternative one based on $SU(2)$. For physical meson masses and decay 
constants the resulting 
Casimir correction turns out to be rather scale-independent and leads to an 
acceptable estimate of the nucleon mass. 
\end{abstract}


\section{Introduction}

It is well-known that the nucleon mass as calculated within the $SU(2)$
Skyrme 
model at tree level comes out to be roughly $50 \%$ too high, 
when empirical values are used for the model parameters. For a long period,
this was considered to be one of the main drawbacks of the model. 
It was only at the beginning of this decade that it was realized\cite{MK91} 
that the proper inclusion of one-loop effects could reduce the skyrmion mass 
down to a value which is in reasonable agreement with the empirical nucleon 
mass. Since then the role of these effects on various nucleon properties have 
been investigated. Quite recently\cite{MW97} it has been shown that, with the 
exception of some axial properties, pionic loop corrections tend to bring 
all the adiabatic quantities close to their experimental values. For those
axial properties there seems to be some difficulties with the $1/N_c$ 
expansion. Still, simple estimates of the next-to-next-to-leading order 
corrections suggest that they could already remedy the defects that show up 
if one includes only adiabatic one-loop contributions.
Given the importance of the loop corrections in several $SU(2)$ skyrmion 
properties, it is clear that their role within the $SU(3)$ Skyrme model
is worth to be investigated. Unfortunately, it is not clear whether the formalism
used in the non-strange sector can be easily extended to the $SU(3)$ model. 
The basic problem is the presence of a rather large symmetry breaking term 
which makes the standard renormalization scheme rather unreliable. In fact, as 
a consequence of such term two alternative approaches to the $SU(3)$
soliton models have been developed. 
One is based on the straighforward extension of the $SU(2)$ 
collective coordinate quantization to the $SU(3)$
flavor group \cite{YA88}.  
Due to the existence of additional rotational modes the predictions
for the tree level baryon masses are even worse than those 
in the $SU(2)$ Skyrme model,
with values which typically lie around $2 \ GeV$. The other approach to strange 
skyrmions\cite{CK85,SNNR88} assumes that the kaon mass is large enough to allow 
only for small amplitude fluctuations along the strangeness direction. 
Thus, hyperons appear as soliton-kaon bound systems. Since in this 
approach there are no additional rotational modes one might expect that 
pion loops 
already account for the necessary Casimir corrections. Then, one has to care about 
the role of the kaon and eta loops which, as we will show later, in the standard 
renormalization scheme diverge in the large mass limit. As we see, even for
the Casimir energies the situation in the $SU(3)$ Skyrme model is unclear and has to 
be carefully studied. As a first step in such investigations one of us recently
reported\cite{WAL97} the results concerning one-loop corrections 
in the flavor symmetric limit $\Delta m = m_K - m_\pi = 0$. 
Such calculations have been done
using the $SU(3)$ collective quantization scheme and show that the
strangeness degrees of freedom tend to push the nucleon mass even
further down. Namely, for the $SU(3)$ case in the flavor symmetric limit the
nucleon mass is $770 \ MeV$ to be compared with the $SU(2)$ prediction 
$1020 \ MeV$. Since these two cases correspond to $\Delta m = 0$ and
$\Delta m \rightarrow \infty$ respectively, for empirical values
of $\Delta m$ the predicted mass is expected to lie between these
two limiting values. However, the methods used in Ref.\cite{WAL97}
cannot be easily used for finite $m_K \neq m_\pi$. In this paper we
introduce an alternative way to evaluate the Casimir correction
which is suitable for large values of $\Delta m$. This will allow us
to compute the one-loop corrections in the bound state approach.

This article is organized as follows. In Sec. 2 we briefly derive
the basic formulae needed to evaluate the one loop corrections to the
skyrmion mass in the bound state approach. In Sec. 3 we discuss
the possible ways to renormalize these corrections. In Sec. 4, we
present the numerical results corresponding to these different
renormalization schemes. Finally, Sec. 5 contains some discussions
and conclusions.  

\section{Formulation}

This investigation is based on the
standard chiral $SU(3)$ lagrangian \cite{gl85} 
\bea\label{lagrangian} 
{\cal L} & = & \frac{F^2}{4} tr \left[ \partial_\mu U \partial^\mu U^\dagger
                  + M ( U + U^\dagger ) \right] \nonumber \\
& + &  (L_1 + L_2 + \frac{1}{2} L_3) ( tr \partial_\mu U \partial^\mu
U^\dagger )^2 + \frac{1}{2} L_2 tr([ U^\dagger \partial_\mu U,
U^\dagger \partial_\nu U ] )^2 \nonumber \\
& + &  (L_3 + 3 L_2 ) \left[ tr \partial_\mu U \partial^\mu U^\dagger 
\partial_\nu U \partial^\nu U^\dagger - \frac{1}{2} 
( tr \partial_\mu U \partial^\mu U^\dagger )^2 \right] \nonumber \\
& + & L_4 tr M ( U + U^\dagger) tr \partial_\mu U \partial^\mu U^\dagger
 +  L_5 tr (U M + M U^\dagger )  \partial_\mu U \partial^\mu U^\dagger
\nonumber \\
& + &  L_6 \left( tr M ( U + U^\dagger) \right)^2 
+ L_7 \left( tr M ( U - U^\dagger ) \right)^2
\nonumber \\
& + & L_8 tr ( M  U M U+ M U^\dagger M U^\dagger ) \nonumber \\
& \equiv & \frac{F^2}{4} tr \left[ \partial_\mu U \partial^\mu U^\dagger
                             + M ( U + U^\dagger ) \right]
+ \sum_{i=1}^8 L_i {\cal L}_i^{(4)} \, .
\eea 
expressed in terms of
the matrix $U$ which contains the dynamical fields and the mass matrix
\be\label{mass}
M = \left( \begin{array}{ccc}
m_\pi^2 & & \\ 
& m_\pi^2 & \\
& & 2m_K^2-m_\pi^2 \\
\end{array} \right) \, .
\ee
In (\ref{lagrangian}) we listed
the familiar non-linear sigma (N$\ell \sigma$) model of
chiral order ChO2 and eight terms of ChO4 which are relevant in the
soliton sector without external fields. The Wess-Zumino-Witten (WZW)
term is included although not explicitly denoted.
After renormalization, the low energy constants (LECs) $L_i(\mu)$
become dependent on the chiral scale $\mu$ and the choice 
$\mu \simeq m_\varrho = 770 MeV$
should provide the lagrangian in leading order $N_c$ \cite{e95}.
Therefore, at this scale
\be\label{zweig}
2 L_1(m_\varrho) - L_2(m_\varrho) =
L_4(m_\varrho) = L_6(m_\varrho) = 0   
\ee
because these combinations are subleading in $N_c$ \cite{gl85}. We
postpone the choice of the remaining LECs to the following section where
we discuss the renormalization schemes.

The starting point of the bound state approach (BSA)  is the hedgehog
solution rotating in $SU(2)$ only,
\be\label{hedgehog}
U_0 = \left( \begin{array}{cc}
A e^{i\mbox{\boldmath$ \tau \cdot \hat r$} F(r)} A^\dagger &  \\ & 1 \\
\end{array} \right) \,\, \qquad A \in SU(2) \, .
\ee
Because the rotation matrix $A$ commutes with the mass matrix
(\ref{mass}) it does not appear in the adiabatic approximation
where time derivatives on the collective coordinates are neglected.
For the description of strange hyperons and also
for the 1-loop calculation considered here, fluctuations $\eta_a$ 
are introduced through the ansatz\cite{CK85}
\be
\label{ansatz}
U = \sqrt{U_0} e^{i \lambda_a \eta_a /F} \sqrt{U_0}  \, ,
\qquad a=1,\dots,8 \, .
\ee
Of particular interest are kaonic ($a=4,\dots,7$) and
eta ($a=8$) fluctuations; pionic fluctuations have already been
calculated in Ref. \cite{MW97}. 
The corresponding equations of motion (e.o.m.) may be generically
written as 
\be \label{eom}
h^2_{ab} \eta_b + i \omega \ {N_c B_0 \over{\sqrt3 f_{\pi}^2}} f_{8ab}\ 
\eta_b = \omega^2 n^2_{ab} \eta_b  \, ,
\ee
where $h^2_{ab}$ is a differential operator, $n^2_{ab}$ the metric and
the linear term in the eigenenergy $\omega$ accounts for the WZW
term that appears in the kaonic case. The partial wave projected
kaonic e.o.m. is explicitely given e.g. in Ref. \cite{SSG95}.
In general, these equations
have to be solved for the phase-shifts. Since they 
decouple for the different meson species into partial waves characterized
by phonon spin $L$ and parity, the pionic, kaonic and eta phase-shifts
may be summed up separately over the various channels $(L, c)$
\be
\delta^x(p)=\sum_{L,c} (2L+1) \delta^x_{L,c}(p) \, \qquad
x = \pi , K , \eta \, .
\ee
The ultra-violet divergencies contained in the Casimir energy
are related to the high momentum behaviour of these phaseshifts
\be
\label{deltas}
\delta^x (p) \stackrel{p \to \infty}{\longrightarrow} a^x_0 p^3 + a^x_1 p +
\frac{a^x_2}{p} + \, \cdots
\ee
with expansion coefficients $a^x_0, a^x_1, a^x_2$ known analytically
for the N$\ell \sigma$ model
(the explicitly denoted terms give rise to at least logarithmically
divergent expressions). 
For the full model (\ref{lagrangian}) the
coefficients have to be determined numerically and the challenge is to
calculate the phase-shifts with great precision up to $p_{max} \simeq
25 m_\pi$ where $L_{max} \simeq 100$ partial waves are needed (for
details see \cite{MW97}). The 1-loop contribution is then
given by \cite{MK91,MW97,WAL97}
\bea
\label{loop}
E_{cas} 
& = & \frac{1}{2 \pi} \sum_x \left\lgroup - \int_0^{\infty}
\frac{p dp}{\sqrt{p^2 + m^2_x}} [\delta^x (p) - a^x_0 p^3 - a^x_1 p - 
\frac{a^x_2}{p}]  - m_x \delta^x (0)  \right. \nonumber\\
&& \qquad + \left. \frac{3 m^4_x a^x_0}{16} (\frac{1}{6} 
+ \ell n \frac{m^2_x}{\mu^2})
 - \frac{m^2_x a^x_1}{4} \ell n \frac{m^2_x}{\mu^2} + \frac{a^x_2}{2}
(1 + \ell n \frac{m^2_x}{\mu^2})  \right\rgroup  \\
&& + \Lambda (\mu) \sum_x \left[ 3 \pi m_x^4 a^x_0 - 4\pi m^2_x a^x_1 + 
8 \pi a^x_2 \right] + {1\over2} \sum_b \omega_b \, . \nonumber 
\eea
The last term in Eq.(\ref{loop}) represents
the contribution coming from all possible bound states. 
The chiral scale $\mu$ is introduced to render the arguments in the logarithms
dimensionless and the divergencies as $d \to 4$ reside in
\be \label{lambda}
\Lambda (\mu) = \frac{\mu^{d-4}}{16 \pi^2} \left[ \frac{1}{d-4} - \frac{1}{2}
(\Gamma'(1) + \ell n (4 \pi) + 1) \right] \, .
\ee
In order to make sense out of these expressions they have to be
properly renormalized. This issue is discussed in the next section.

\section{The renormalization schemes}

There are various possibilities to renormalize the expression
(\ref{loop}) which we are going to discuss subsequently.

\subsection{Standard $SU(3)$ renormalization scheme}

Noting that the N$\ell \sigma$ model expansion coefficients obey the
ChO4 relation
\be
\sum_x \left[ 3 \pi m_x^4 a^x_0 - 4\pi m^2_x a^x_1 + 8 \pi a^x_2 \right]
= \sum^8_{i=1} \Gamma_i \int \, d^3 r {\cal L}_i^{(4)} \, ,
\ee
the divergencies in (\ref{loop}) may be absorbed into a redefinition
of the lagrangian's LECs
\be \label{lec}
L_i (\mu) = L_i - \Gamma_i \Lambda (\mu) \, , \qquad
L_i(\mu)=L_i (m_{\varrho})-\frac{\Gamma_i}{32\pi^2}
\ell n(\frac{\mu^2}{m_{\varrho}^2}) 
\ee
which become scale-dependent. The
$\Gamma_i$'s are simple
numerical factors given in Table 1 and coincide with those
defined in \cite{gl85}. $F^2$ and the mass matrix $M$ are 
not renormalized in that scheme
which is identical to that used in $SU(3)$ chiral perturbation theory.

The remaining piece in (\ref{loop}) is the finite Casimir energy for
the three meson species
\bea
\label{casimir}
E^x_{\mbox{\scriptsize\ cas}} (\mu) 
& = & - \frac{1}{2 \pi} \int_0^{\infty}
\frac{p dp}{\sqrt{p^2 + m^2_x}} [\delta^x (p) - a^x_0 p^3 - a^x_1 p - 
\frac{a^x_2}{p}]  - \frac{m_x}{2 \pi} \delta^x (0)  \\
&& + \frac{3 m^4_x a^x_0}{32 \pi} (\frac{1}{6} 
+ \ell n \frac{m^2_x}{\mu^2})
 - \frac{m^2_x a^x_1}{8 \pi} \ell n \frac{m^2_x}{\mu^2} +
\frac{a^x_2}{4 \pi}
(1 + \ell n \frac{m^2_x}{\mu^2})  
+ {1\over2} \sum_b \omega_b \nonumber
\eea
which has to be added to the static soliton mass calculated with the 
renormalized LECs (\ref{lec})
\be
E_{\mbox{\scriptsize\ tree+1-loop}} 
= M_{\mbox{\scriptsize\ sol}} (\mu) + 
E^\pi_{\mbox{\scriptsize\ cas}} (\mu) + 
E^K_{\mbox{\scriptsize\ cas}} (\mu) +
E^\eta_{\mbox{\scriptsize\ cas}} (\mu) 
\, .
\ee
In order to obtain at scale $\mu=m_\varrho$ the same soliton as in the
$SU(2)$ calculation \cite{MW97} we fix $F=91.1 MeV$ and the remaining
LECs according to (\ref{zweig}) and
\bea\label{b}
&& L_2(m_\varrho) =  \frac{1}{16e^2} \, , \qquad
L_3(m_\varrho)+3 L_2(m_\varrho) = 0 \, 
\nonumber \\
&& L_5(m_\varrho) = 2L_8(m_\varrho) = -6L_7(m_\varrho) = 2.3 \cdot 10^{-3}  
\eea 
which is in accordance with the standard values \cite{gl85,e95}
(within error bars) with one exception: $L_2$ has to be replaced by an
effective Skyrme parameter $e=4.25$ (the standard value would
correspond to $e \simeq 7$) in order to simulate the missing higher
ChOs generated by vector mesons. 
A detailed justification of this
choice used also in \cite{WAL97} is found in Ref.  \cite{MW97}.
It should be mentioned that although
the LECs are chosen such that many of the ChO4 terms in
(\ref{lagrangian}) vanish at scale $\mu = m_\varrho$ all these terms are
switched on and do contribute when the scale is changed.

As it is obvious from Eq.(\ref{casimir}), this renormalization scheme
becomes increasingly unreliable as $m_x$ becomes closer or larger
than the renormalization scale $\mu$. In such cases, an alternative
scheme is needed.

\subsection{$SU(2)$ renormalization scheme}

For large kaon mass the troublesome terms are located in the kaon's
and eta's Casimir energies. While the phase-shift integral 
in Eq.(\ref{casimir}) behaves well, the extra terms multiplied with the
chiral logs explode with increasing kaon mass. It is now possible
to renormalize also these terms into the lagrangian
\be\label{lagbar} 
\bar {\cal L} = 
\frac{\bar F^2}{4} tr \left[ \partial_\mu U \partial^\mu U^\dagger
                             + \bar M ( U + U^\dagger ) \right]
+ \sum_{i=1}^8 \bar L_i {\cal L}_i^{(4)} \, .
\ee 
with the $\bar L^{\prime}_i$s given in Table 1 and
\bea\label{bar} 
\bar F^2&=&F^2 (1-2 \mu_K) + m_\pi^2 \nu_K \nonumber \\
\bar F^2 \bar m_x^2&=&m_x^2 \left[ F^2 (1-2 \mu_K-\frac{1}{3} \mu_\eta) 
+ m_\pi^2 ( \nu_K + \frac{1}{9} \nu_\eta) - (2m_K^2-m_\pi^2)
(\frac{1}{6} \nu_K + \frac{1}{9} \nu_\eta) \right] \nonumber \\ 
\mbox{where} &&
\nu_x = \frac{1}{32 \pi^2}(1 + \ell n \frac{m^2_x}{\mu^2}) \, , \qquad  
\mu_x = \frac{m_x^2}{32 \pi^2 F^2} \ell n \frac{m^2_x}{\mu^2} \, .  
\eea 
All these quantities become scale-dependent. For $\bar F$ and 
$\bar m_x$ this dependence may be read directly from 
their definition (\ref{bar}) and
correspondingly for the $\bar L_i$
\be \label{lecbar}
\bar L_i(\mu)=\bar L_i (m_{\varrho})-\frac{\bar \Gamma_i}{32\pi^2}
\ell n(\frac{\mu^2}{m_{\varrho}^2}) 
\ee
with coefficients $\bar \Gamma_i$ obtained in $SU(2)$ 
\cite{gl85} and listed in Table 1.
However, in contrast to the standard $SU(3)$ coefficients, the bared LECs
$\bar L_i$ do not depend on the kaon mass; in fact the kaon mass
dependence of the $SU(3)$ LECs $L_i$ may be deduced from these
relations.

In order to make the connection to $SU(2)$ more transparent we give the
relation for the corresponding LECs $f$, $m$, and $\ell_i (i=1,\dots,4)$
explicitely \cite{gl85}
\bea\label{su2} 
&& f^2 = \bar F^2 + 8(2m_K^2-m_\pi^2) \bar L_4 \nonumber \\
&& f^2 m^2 =\bar m_\pi^2 [ \bar F^2 +16(2m_K^2-m_\pi^2) \bar L_6] \nonumber \\
&& \ell_1 = 4 \bar L_1 + 2 \bar L_3 \nonumber \\
&& \ell_2 = 4 \bar L_2 \nonumber \\
&& \ell_3 = -8 \bar L_4 - 4 \bar L_5 
    + 16 \bar L_6 + 8 \bar L_8 \nonumber \\
&& \ell_4 = 8 \bar L_4 + 4 \bar L_5   \, .
\eea 
It is readily checked that the $\ell_i$ have the correct scale-dependence
\cite{gl84}.

Let us turn back to the 1-loop calculation of the soliton energy. While
the expression for the pionic Casimir energy is unaltered the kaonic
and eta Casimir energies are now given exclusively by the phase-shift
integral and the bound state contributions
\bea\label{casbar}
\bar E^x_{\mbox{\scriptsize\ cas}} (\mu) 
&=&  - \frac{1}{2 \pi} \int_0^{\infty}
\frac{p dp}{\sqrt{p^2 + m^2_x}} [\delta^x (p) - a^x_0 p^3 - a^x_1 p - 
\frac{a^x_2}{p}]  - \frac{m_x}{2 \pi} \delta^x (0) \nonumber \\ 
&& +  {1\over2} \sum_b \omega_b \, ,
\qquad \qquad x=K,\eta
\eea
without the troublesome extra terms. Therefore the total soliton energy
in tree + 1-loop is given by
\be
\bar E_{\mbox{\scriptsize\ tree+1-loop}} 
= \bar M_{\mbox{\scriptsize\ sol}} (\mu) + 
E^\pi_{\mbox{\scriptsize\ cas}} (\mu) + 
\bar E^K_{\mbox{\scriptsize\ cas}} (\mu) +
\bar E^\eta_{\mbox{\scriptsize\ cas}} (\mu)  \, .
\ee
Note that for kaons the last two terms in Eq.(\ref{casbar}) 
exactly cancel out in the limit of infinite kaon mass. As a consequence
the kaonic contribution remains finite even in that limit.

\section{Numerical results}

The physical kaon and eta masses are somewhat smaller, although comparable, 
than the chosen scale $m_\rho$. In this sense, it is not a priori
clear which renormalization scheme is more appropiate. We investigate
this issue numerically. We consider first the $SU(3)$ renormalization scheme
assuming that the meson masses contained in the mass matrix take their physical 
values. 
As a consequence of the choice (\ref{b}) the pionic Casimir energy is the
same as in the $SU(2)$ case where we found scale-independence 
of the soliton energy in tree +
1-loop over a wide region of chiral scales \cite{MW97}. It should be
mentioned that the scale-dependence of the LECs (\ref{lec}) differs
slightly from that of the LECs used in $SU(2)$ but that does not
affect the statement concerning the scale-independence of the
soliton energy.
Because the coupling of the eta to the soliton proceeds through the
mass terms only, it is extremely weak:
the Casimir energy of the eta amounts to
$E^\eta_{\mbox{\scriptsize\ cas}} (m_\rho) = 1.7 MeV$ and can savely be
omitted. Therefore in what follows we concentrate on the kaonic 
contribution to the Casimir
energy  $E^K_{\mbox{\scriptsize\ cas}}$. In this sector we get
two doublets of bound states. For the model parameters we use, the
S-wave doublet appears at $\omega_S = 465 \ MeV$ and the P-doublet
at $\omega_P = 256 \ MeV$.

The kaonic Casimir energy is shown in Fig. 1 depending on the chiral
scale $\mu$ (dashed line). 
For $\mu=m_\rho=770 MeV$ we obtain a positive contribution
of $\simeq 140 MeV$ which has to be added to the $SU(2)$ soliton energy
in tree + 1-loop of $1020 MeV$ in order to get the total nucleon
mass. We observe a considerable scale-dependence of the kaonic
Casimir energy which may have one of the following reasons: (i)
there may be important terms missing in the lagrangian, (ii) the 
hedgehog rotating in $SU(2)$ as used in the BSA
might be not an appropriate solution for physical kaon mass, or (iii)
the employed renormalization scheme could be incompatible with that
approach. 

In order to shed some light onto this issue we are going to
investigate the dependence on the kaon mass in more detail.
Particularly, we
know that the BSA should become accurate for large kaon masses.
The dependence of the kaonic Casimir energy on the kaon mass is
shown in Fig. 2 (dashed line). We observe a drastic dependence and,
as expected, we notice that the kaonic Casimir energy does not
tend to zero for large kaon masses as it should. This is of course no 
surprise because the experimental LECs (\ref{lec}) were evaluated for 
physical kaon mass and should be used only there. This study of the kaon mass
dependence indicates that the $SU(2)$ renormalization scheme which allows
the LECs to run smoothly into their $SU(2)$ values for large kaon masses
might be more suitable for the BSA.

In the $SU(2)$ renormalization scheme we start by choosing
at the scale $\mu= m_\rho$ the bared quantities in exactly the
same way as the unbared ones above. This guarantees that the soliton
mass and the pionic Casimir energy as well as their 
scale-dependences are exactly the same as 
in $SU(2)$. Note that this was not exactly true in 
the $SU(3)$ renormalization scheme where the LECs scaled 
in a slightly different way. We may again forget about the tiny 
eta Casimir energy 
($\bar E^\eta_{\mbox{\scriptsize\ cas}} (m_\rho)=0.006 MeV$) and
concentrate on the kaonic Casimir energy 
$\bar E^K_{\mbox{\scriptsize\ cas}}$ only.

As before we consider first the physical value for the kaon mass. Compared to
$E^K_{\mbox{\scriptsize\ cas}}$ in the $SU(3)$ renormalization scheme
the scale-dependence has reduced appreciably (Fig. 1, full line)
indicating that indeed the $SU(2)$ renormalization scheme is more appropriate
for the BSA. At scale $\mu=m_\rho$ we obtain
$\bar E^K_{\mbox{\scriptsize\ cas}}(m_\rho)=-140 MeV$ which added to
the $SU(2)$ value of $1020 MeV$ yields an acceptable estimate for the
nucleon mass. As is noticed from Fig. 2 (full line) the dependence on
the kaon mass seems to be reasonable, as expected i.e. the kaonic
Casimir energy tends to zero with increasing kaon mass.

\section{Discussion and conclusions}

We have studied the one-loop corrections to the $SU(3)$ skyrmion mass 
within the bound state soliton model. This approach assumes
that along the strangeness direction only small amplitude 
fluctuations are possible. Thus, it is expected to become
more accurate as the $SU(3)$ symmetry breaking terms increase.
We found that the standard $SU(3)$ renormalization scheme
leads to a considerable scale-dependence of the correction. Therefore,
we introduced an alternative $SU(2)$ renormalization scheme.
Such scheme yields an estimate of the total skyrmion mass
which is in reasonable agreement with the empirical nucleon
mass. Of course, this procedure is exact only in the limit $m_K=\infty$.
To choose $m_K$ large but finite is an approximation since 
as soon as $m_K$ is finite (i) the classical solution should start to explore 
the strange sector and (ii) the $SU(2)$ renormalization 
scheme will be modified. 
Nevertheless, the BSA may be a good approximation for sufficiently large $m_K$.
Actually, one could argue that this method might be trusted down to the region 
where the solid curve in Fig.2 has its minimum. For smaller $m_K$ the
kaonic Casimir energy 
increases again although we know that the flavor symmetric value 
(dot in Fig. 3) lies much 
lower. Still, if one makes a smooth interpolation from the 
minimum up to the flavor 
symmetric point the resulting kaonic contribution does not differ very much
from the value quoted above, $\bar E^{m_K}_{cas} (m_\rho) = -140 MeV$. 
In any case, what seems to be clear from Fig.2 is that the standard 
$SU(3)$ scheme together
with BSA cannot be used to perform such interpolation.
The slow rotator approach \cite{SW92} appears as much well suited for that.
However already in that approach the eta and kaonic 
components which are driven 
by the isospin breaking are missing. Thus, not even the slope of the 
Casimir energy at $m_K=m_\pi$ can be easily exactly calculated. Nevertheless, the 
simpler rigid rotator approach may still be a good approximation for sufficiently 
small $m_K > m_\pi$. 
Therefore, it would be interesting to complement the results of the present
work with a calculation of the Casimir energy within the rigid rotator approach.
There is even a possibility to obtain a smooth transition from small to large kaon 
masses in that way.

N.N.S. is fellow of the CONICET, Argentina. He also acknowledges a
grant of the Fundaci\'on Antorchas, Argentina. 
H.W. is supported by a grant
of JNICT, Portugal (Contract PRAXIS/2/2.1/FIS/451/94).

\vspace*{0.5cm}

{\bf Note added:}

\vspace*{0.5cm}

Having finished this manuscript we became aware of a recently published paper
by Kim and Park \cite{KP98}, which also reports on the kaonic Casimir
energy. Using Moussallam's formula and neglecting the counter terms,
their approach essentially corresponds to the standard $SU(3)$
regularization scheme discussed in Section 3.1. The difference of their
number $+100 MeV$ versus our $+140 MeV$ for that case is mainly
because we included all ChO4 terms in our lagrangian, most prominantly
the kinetic symmetry breaker which influences the bound-state energies
considerably. However, from our Fig.2 it becomes obvious that the
kaonic Casimir energy should be negative in order to interpolate
between $SU(2)$ and the $SU(3)$ flavor symmetric result. For that 
reason we considered the $SU(2)$ renormalization scheme (Section 3.2)
more appropriate. This choice was supported by the moderate 
scale-dependence found in that case (Fig.1).

\vspace*{2cm}

\begin{table}[h]
\begin{center}
\caption{
{\it Relation between the LECs $\bar L_i$ and the standard LECs
$L_i$ together with the corresponding $SU(2)$ and $SU(3)$ coefficients 
$\bar \Gamma_i$ and $\Gamma_i$; $\nu_K$ and $\nu_\eta$} are defined
in Eq.(\ref{bar}).
}

\vspace*{1.cm} 
\begin{tabular}{|lll|}
\hline
&&\\
$\qquad \bar L_1 = L_1 - \frac{1}{96} \nu_K$ 
    & $\Gamma_1 = \frac{3}{32}$ & $\bar \Gamma_1 = \frac{1}{12} \qquad$ \\
&&\\
$\qquad \bar L_2 = L_2 - \frac{1}{48} \nu_K$ 
    & $\Gamma_2 = \frac{3}{16}$ & $\bar \Gamma_2 = \frac{1}{6}$  \\ 
&&\\
$\qquad \bar L_3 = L_3$  
    & $\Gamma_3 = 0 $ & $\bar \Gamma_3 = 0 $ \\ 
&&\\
$\qquad \bar L_4 = L_4$  
    & $\Gamma_4 = \frac{1}{8} $ & $\bar \Gamma_4 = \frac{1}{8}$  \\ 
&&\\
$\qquad \bar L_5 = L_5 - \frac{1}{8} \nu_K$ 
    & $\Gamma_1 = \frac{3}{8} $ & $\bar \Gamma_5 = \frac{1}{4}$  \\ 
&&\\
$\qquad \bar L_6 = L_6 + \frac{1}{96} \nu_K + \frac{1}{144} \nu_\eta $ 
    & $\Gamma_6 = \frac{11}{144} \qquad $ & $\bar \Gamma_6 = \frac{3}{32}$  \\ 
&&\\
$\qquad \bar L_7 = L_7$  
    & $\Gamma_7 = 0 $ & $ \bar \Gamma_7 = 0 $ \\ 
&&\\
$\qquad \bar L_8 = L_8 - \frac{1}{12} \nu_K - \frac{1}{48} \nu_\eta
\qquad \qquad$ 
    & $\Gamma_8 = \frac{5}{48}$ & $\bar \Gamma_8 = 0$  \\ 
&&\\
\hline
\end{tabular}
\end{center} 
\end{table}

\vfill

\pagebreak

                 
                 

\begin{figure}
\centerline{\psfig{figure=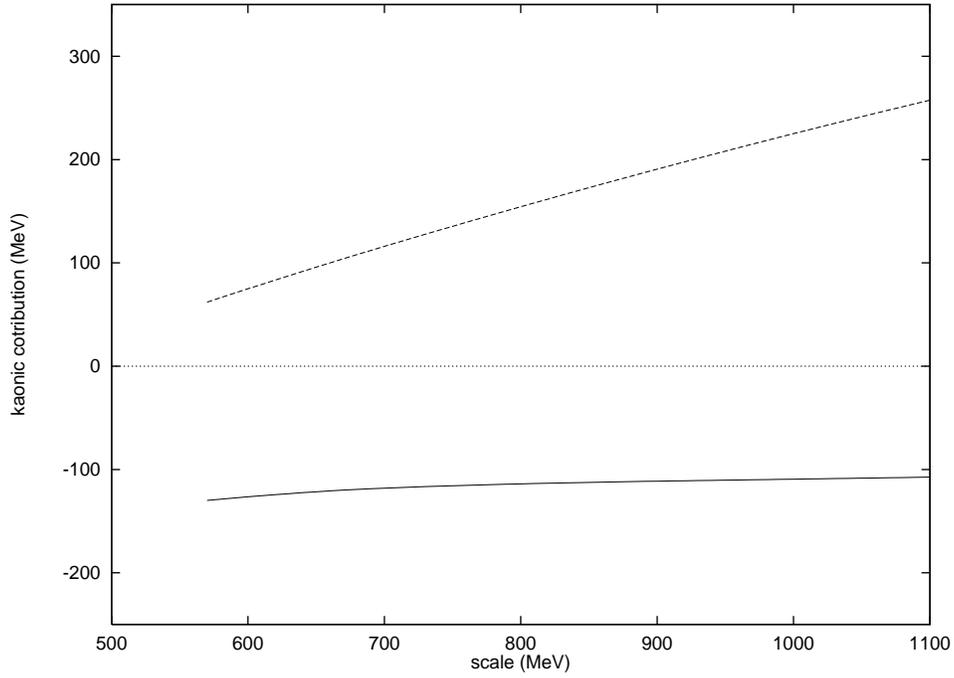,height=9.cm,angle=-90}}
\vspace{1.cm}
\protect\caption{\it Scale dependence of the Casimir corrections
for empirical meson masses. Dashed line corresponds to the $SU(3)$
renormalization scheme whereas the full line corresponds to the $SU(2)$
scheme.}
\label{scaledep}
\end{figure}

\begin{figure}
\centerline{\psfig{figure=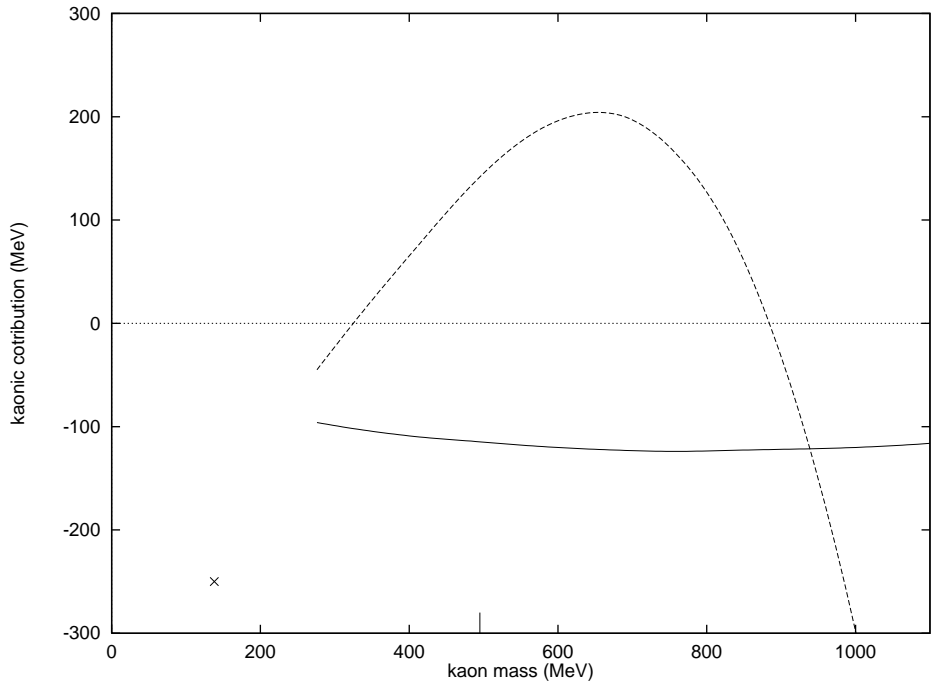,height=9.cm,angle=-90}}
\vspace{1.cm}
\protect\caption[]{\it Kaon mass dependence of the Casimir corrections.
Dashed line corresponds to the $SU(3)$
renormalization scheme whereas the full line corresponds to the $SU(2)$
scheme. The cross represents the flavor symmetric limit considered
in Ref.\cite{WAL97}.
The tick at $495 MeV$ indicates the empirical kaon mass.}
\label{massdep}
\end{figure}


\begin{thebibliography}{80}

\bibitem{MK91}    B. Moussallam and D. Kalafatis,
                  Phys. Lett. {\bf B272}, 196 (1991);
                  B. Moussallam, 
                  Ann. Phys. {\bf 225}, 264 (1993).

\bibitem{MW97}    F. Meier and H. Walliser, 
                  Phys.Rep. {\bf 289}, 383 (1997). 

\bibitem{YA88}    H. Yabu and K. Ando,
                  Nucl. Phys. {\bf B301}, 601 (1988).

\bibitem{CK85}    C.G. Callan and I. Klebanov,
                  Nucl. Phys. {\bf B262}, 365 (1985).

\bibitem{SNNR88}  N.N. Scoccola, H. Nadeau, M. Nowak and M. Rho,
                  Phys. Lett. {\bf B201}, 425 (1988);
                  C.G. Callan, K Hornbostel and I. Klebanov, 
                  Phys. Lett. {\bf B202}, 269 (1988);
                  U. Blom, K. Dannbom and D.O. Riska,
                  Nucl. Phys. {\bf A493}, 384 (1989).

\bibitem{WAL97}   H. Walliser, hep-ph/9710232 
                  (Phys. Lett. {\bf B}, in press).

\bibitem{gl85}    J. Gasser and H. Leutwyler, 
                  Nucl. Phys. {\bf B250}, 465 (1985).

\bibitem{e95}     G. Ecker, 
                  Czech. J. of Phys. {\bf 44}, 405 (1995).

\bibitem{SSG95}   C. Schat, N.N. Scoccola and C. Gobbi, 
                  Nucl. Phys. {\bf A585}, 627 (1995). 

\bibitem{gl84}    J. Gasser and H. Leutwyler, 
                  Ann. Phys. {\bf 158}, 142 (1984).

\bibitem{SW92}    B. Schwesinger and H. Weigel,
                  Nucl. Phys. {\bf A540}, 461 (1992).

\bibitem{KP98}    J.-I. Kim and B.-Y. Park,
                  Phys. Rev. {\bf D57}, 2853 (1998).

\end{thebibliography}
\end{document}